\documentclass[letterpaper, 10pt, conference]{ieeeconf}  

\IEEEoverridecommandlockouts                              

\usepackage{float}
\usepackage{graphicx}

\makeatletter
\let\NAT@parse\undefined
\makeatother
\usepackage[numbers,sort&compress,square]{natbib}

\usepackage{amsmath} 
\usepackage{amssymb}  
\usepackage{amsfonts}  

\title{\LARGE \bf
Control of a Floating Wind Turbine on a Novel Actuated Platform
}

\author{David Stockhouse*$^{1}$, Mandar Phadnis$^{1}$, Elenya Grant$^{2}$,
Kathryn Johnson$^{2}$, Rick Damiani$^{2}$, and Lucy Pao$^{1}$
\thanks{This work was supported, in part, by the Advanced Research Projects
Agency Energy (ARPA-e) Aerodynamic Turbines Lighter and Afloat with Nautical
Technologies and Integrated Servo-control (ATLANTIS) Program, under Award Number
DE-AR0001181. Any opinions, findings, and conclusions or recommendations
expressed in this material are those of the authors and do not necessarily
reflect the views of ARPA-e.}%
\thanks{The authors would like to thank Senu Sirnivas and the National Renewable
Energy Laboratory (NREL) for model creation and OpenFAST development support. We
also thank the entire USFLOWT team for collaborative development, and Manuel
Pusch for revision suggestions.}%
\thanks{*Corresponding author. {\tt David.Stockhouse@colorado.edu}}%
\thanks{$^{1}$University of Colorado Boulder, Boulder, CO}%
\thanks{$^{2}$Colorado School of Mines, Golden, CO}%
}

\begin{document}


\onecolumn

\hspace{0pt}
\vfill

\begin{centering}

%

This work has been submitted to the IEEE for possible publication. Copyright may be transferred without notice, after which this version may no longer be accessible.

\end{centering}

\vfill
\hspace{0pt}

\twocolumn

\newpage

\maketitle
\thispagestyle{empty}
\pagestyle{empty}

\begin{abstract}

Designing a floating offshore wind turbine (FOWT) controller requires solving engineering challenges not found for fixed-bottom turbines. This paper applies several methods from the growing body of FOWT control literature to the 10-MW Ultraflexible Smart FLoating Offshore Wind Turbine (USFLOWT) baseline generator-speed controller. USFLOWT aims to reduce capital expenses using the lightweight SpiderFLOAT platform, a novel smart floating substructure with built-in distributed actuators for direct platform tilt and heave control. In this work, the USFLOWT baseline controller is improved through detuning and parallel compensation with both blade pitch and generator torque. The SpiderFLOAT platform additionally allows motion compensation through distributed platform actuators. Two proposed SpiderFLOAT actuator types are considered for active platform control: a low-bandwidth actuator that uses variable floater ballast to bring a heeling platform to a steady-state upright position, and a high-bandwidth actuator that dynamically changes the substructure geometry to actively reject transient platform motion.  Each control approach is tested for USFLOWT using the open-source aero-hydro-servo-elastic wind turbine simulation tool OpenFAST.  Performance results for each approach are compared across a range of above-rated wind speeds, and promising combined approaches are further evaluated to recommend future multi-parameter optimization pathways.

\end{abstract}


\section{INTRODUCTION}

The USFLOWT project aims to reduce levelized cost of energy (LCOE) of floating
wind turbines by using smart turbine and platform control to increase power
while mitigating structural loads. Two approaches in the literature that have
matured in recent years are detuning~\cite{larsen2007} and parallel
compensation~\cite{vanderveen2012,fischer2013}, the latter of which increases
the complexity of the controller while significantly improving
performance~\cite{fleming2016,yu2018}. A novel advantage of USFLOWT that aids in
the control challenge is the flexible SpiderFLOAT~\cite{damiani2021} platform
with actuators for direct platform tilt control. 

A fundamental problem with FOWT control is generator speed instability, brought
about by large feedback gains in the presence of non-minimum phase zeros (NMPZs)
due to the platform fore-aft mode.  This non-minimum phase characteristic can
also be found for fixed-bottom and land-based turbines, though the platform
frequency of a FOWT is generally much lower than the tower fore-aft frequency of
a fixed-bottom turbine. Correspondingly, the NMPZs of a FOWT system can lead to
an unstable platform resonance mode~\cite{larsen2007}. The destabilizing effect
of NMPZs has been referred to as ``negative damping" in the FOWT
literature~\mbox{\cite{larsen2007,jonkman2008,vanderveen2012,fischer2013,fleming2016}},
but in this paper it is simply called instability. Detuning the blade pitch
controller to avoid exciting the platform mode sacrifices responsiveness to wind
disturbances and risks generator overspeeding that may trigger turbine
shutdowns. Thus, detuning alone does not provide a complete solution to the NMPZ
problem, and FOWT-specific control techniques are required.

Another method of mitigating the NMPZ effects involves leveraging the
conventional blade pitch~\cite{vanderveen2012} and generator
torque~\cite{fischer2013} actuators in an independent compensation loop to
reduce the coupling of platform motion to generator speed. This parallel
compensation scheme is effective, but the turbine actuators suffer from limited
control authority over platform motion while being primarily employed for
generator speed regulation and maximizing power production. Because of the tight
coupling to power generation, utilizing these actuators for fore-aft motion
compensation is subject to inherent tradeoffs.

The novelty of USFLOWT is the introduction of platform actuators which can
directly influence platform degrees of freedom (DOFs) without direct coupling to
generator speed control. Utilizing these actuators to compensate for platform
motion brings the system into the domain of multi-input multi-output (MIMO)
control. The SpiderFLOAT actuators on their own do not significantly increase
power capture, but through load mitigation they facilitate LCOE reduction made
by other control and design parameters. Control co-design, another growing
research area within the wind turbine \mbox{industry~\cite{garcia-sanz2019}}, offers
the ability to optimize power capture through iterative control and design
parameter modifications, and also allows secondary control objectives like
structural load reduction to further reduce LCOE in a more optimal solution than
uncoupled individual optimizations can achieve.

This paper is organized as follows. In Section~\ref{sec:wind_turbine}, the wind
turbine and floating platform used for this study are presented along with a
low-order linear model. Section~\ref{sec:control_design} details the design and
tuning of each control feature, starting from tuning of the land-based
controller and leading to detuning for improving stability, to parallel
compensation, and finally utilizing the SpiderFLOAT actuators in the complete
FOWT controller. The performance of these controllers simulated in OpenFAST is
presented in Section~\ref{sec:simulation_results}, and conclusions are
discussed in Section~\ref{sec:conclusions}.

\section{WIND TURBINE DESCRIPTION} \label{sec:wind_turbine}

\begin{figure}[t]
    \centering
    \includegraphics[scale=.5]{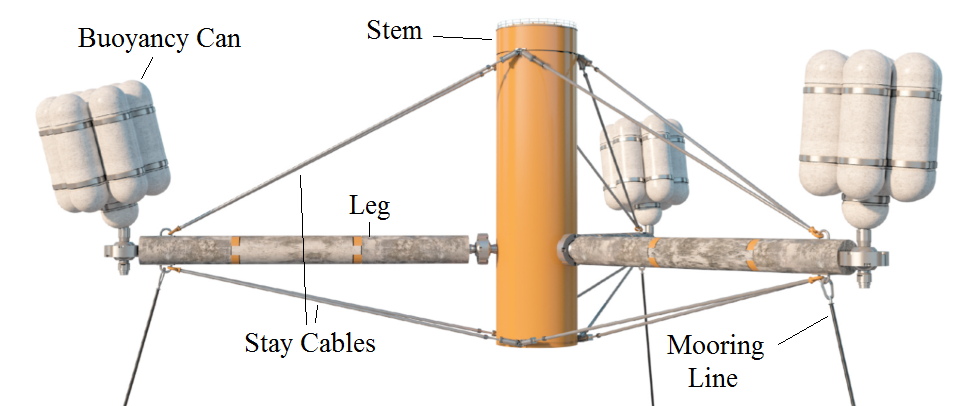}
    \vspace*{-0.75cm}
    \caption{The SpiderFLOAT platform \cite{damiani2021}[Image credit: J. Bauer, NREL].}
    \label{fig:sf}
    \vspace*{-0.1cm}
\end{figure}

The FOWT system used in this study is the \mbox{USFLOWT}, composed of the DTU-10MW
reference wind turbine~\cite{dtu10} supported by a novel bio-inspired
substructure called \mbox{SpiderFLOAT~\cite{damiani2021}}.  The DTU-10MW has rated and
cut-out wind speeds of 11.4 m/s and 25 m/s, respectively. The controllers
analyzed in this paper are used at wind speeds between rated and cut-out
(region 3). For an overview of conventional fixed-bottom wind turbine control
design, see~\cite{pao2011}. The SpiderFLOAT substructure is visualized in
Fig.~\ref{fig:sf}, in which the buoyancy cans have controllable ballast and the
stay cables have controllable length.

\subsection{FOWT Model} \label{sec:modeling}

The dynamics of immediate interest for FOWT controller design are those of the
generator and the platform pitch. The fourth-order wind turbine model
construction followed here is described in detail in~\cite{abbas2021}, with the
most relevant aspects summarized here. A first-order model of a wind turbine
describes the dynamics of the generator speed, $\omega_g$, with respect to the
aerodynamic torque, $\tau_a$, and the generator resistance torque, $\tau_g$:
\begin{equation} \label{eq:rotor}
    \dot{\omega}_g = \frac{N_g}{J_r} \left(\tau_a(\omega_g, v, \beta) - N_g \tau_g \right),
\end{equation}
where $\beta$ is the blade pitch angle, $v$ is the rotor averaged wind speed,
$N_g$ is the gearbox ratio, and $J_r$ is the rotor inertia. The aerodynamic
torque is a nonlinear function of rotor variables ($\omega_g$, $v$, and $\beta$)
described numerically by rotor performance surfaces calculated by a rotor
aerodynamic solver.  Stationary solutions with $\overline{\dot{\omega}_g} = 0$,
$\overline{\omega_g} = \omega_{g,rated}$, $\overline{\tau_g} = \tau_{g,rated}$,
and a corresponding $\overline{\beta}$ exist at each above-rated wind speed
operating point $\overline{v}$, and we can numerically compute a linearized
model describing the system response to small perturbations in any of the state
variables at each operating point.  For notation, the system dynamics at a
particular operating point are linearized around the steady-state variables
$\overline{x}$, and the linearized dynamics are with respect to perturbations
$\tilde{x}$ relative to $\overline{x}$. The value of each variable in absolute
terms can be reconstructed as $x = \overline{x} + \tilde{x}$. The linearized
form of Eq.~\eqref{eq:rotor} is
\begin{equation} \label{eq:rotor_lin}
    \tilde{\dot{\omega}}_g = \frac{N_g}{J_r} \left( \frac{\partial \tau_a}
    {\partial \omega_g} \tilde{\omega}_g + 
        \frac{\partial \tau_a}{\partial v} \tilde{v} + 
        \frac{\partial \tau_a}{\partial \beta} \tilde{\beta} \right) -
        \frac{N_g^2}{J_r} \tilde{\tau}_g,
\end{equation}
where each partial derivative is evaluated at $\overline{\omega_g}$,
$\overline{v}$, and $\overline{\beta}$. These are estimated numerically from the
rotor performance surfaces for each operating point.

Tuning FOWT-specific control features requires a model that incorporates the
most critical floating platform dynamics. In this model, the dynamic response of
the platform pitch $\phi$ is approximated as a second-order oscillator
aggregating the damping and restoring forces provided by the platform and
mooring lines, and the other platform DOFs are ignored. The pitch direction is
defined such that $\phi>0$ means that the platform is tilting downwind and
assumed small enough that the small-angle approximation applies. The
second-order platform dynamics are represented by
\begin{equation} \label{eq:ptfm}
    J_t \ddot{\phi} + D_t \dot{\phi} + k_t \phi = h_t F_a(\omega_g, v_r, \beta)
    + \tau_p,
\end{equation}
where $J_t$ is the total system moment of inertia about the pitch rotational
mode, $D_t$ is the natural damping coefficient (assumed constant), $k_t$ is a
spring-like restoring coefficient, $h_t$ is the height of the rotor
(approximately tower length), $\tau_p$ is the torque supplied by the platform
actuators, and $F_a$ is the aerodynamic thrust at the rotor, a nonlinear
function of the rotor variables.  Platform pitch motion is experienced by the
rotor as relative wind, so $v_r = v - h_t \dot{\phi}$.  This relative wind speed
also influences the aerodynamic torque equations.

The simplified model in~\eqref{eq:ptfm} ignores many higher-order dynamics of
the USFLOWT system, including additional platform DOFs (such as surge and
heave), mooring dynamics, and the complex flexibility and hydrodynamic response
of the SpiderFLOAT platform. This low-order model is sufficiently detailed for
an automated tuning process if the second-order response can be identified from
simulation data and the platform does not experience extreme displacement during
operation.  Similar to Eq.~\eqref{eq:rotor_lin}, the thrust force $F_a$ can be
linearized at a given operating point:
\begin{align} \label{eq:ptfm_lin}
    J_t \tilde{\ddot{\phi}} + &\left(D_t + h_t \frac{\partial F_a}{\partial v} \right) \tilde{\dot{\phi}}
    + k_t \tilde{\phi} = \\ \nonumber
       &h_t \left( \frac{\partial F_a}{\partial \omega_g} \tilde{\omega}_g + 
        \frac{\partial F_a}{\partial v} \tilde{v} + 
        \frac{\partial F_a}{\partial \beta} \tilde{\beta} \right) +
        \tilde{\tau}_p.
\end{align}
The simplified platform actuator model used here ignores most of the dynamics
within the SpiderFLOAT substructure. It is abstracted as a single applied
moment, $\tau_p$, whose impact on the system is identified using empirical
simulation data.

Combining the linear models for generator dynamics and platform pitch yields a
state-space model with state $\boldsymbol{x}~=~\begin{bmatrix} \tilde{\theta} &
\tilde{\dot{\theta}} & \tilde{\phi} & \tilde{\dot{\phi}} \end{bmatrix}^\intercal$ and
input $\boldsymbol{u}~=~\begin{bmatrix} \tilde{v} & \tilde{\beta} &
\tilde{\tau}_g & \tilde{\tau}_p \end{bmatrix}^\intercal$, where $\theta$ is the
integral of generator speed used for integral control, defined so that
$\dot{\theta} = \omega_g$ and $\tilde{\theta} = \int{\tilde{\omega}_g dt}$. The
combined linearized state-space model is $\boldsymbol{\dot{x}} = \boldsymbol{A} \boldsymbol{x} + \boldsymbol{B} \boldsymbol{u}$:
\begin{align} \label{eq:ss_model}
    \boldsymbol{\dot{x}} = &\begin{bmatrix}
        0 & 1 & 0 & 0 \\
        0 & \frac{N_g}{J_r} \frac{\partial \tau_a}{\partial \omega_g} & 0 & -h_t \frac{N_g}{J_r} \frac{\partial \tau_a}{\partial v} \\
        0 & 0 & 0 & 1 \\
        0 & \frac{h_t}{J_t} \frac{\partial F_a}{\partial \omega_g} & \frac{-k_t}{J_t} & \frac{-1}{J_t} \left(D_t + h_t \frac{\partial F_a}{\partial v} \right)
    \end{bmatrix} \boldsymbol{x} \\ 
    & + \begin{bmatrix}
        0 & 0 & 0 & 0 \\
        \frac{N_g}{J_r} \frac{\partial \tau_a}{\partial v} & \frac{N_g}{J_r} \frac{\partial \tau_a}{\partial \beta} & -\frac{N_g^2}{J_r} & 0\\
        0 & 0 & 0 & 0 \\
        \frac{h_t}{J_t} \frac{\partial F_a}{\partial v} & \frac{h_t}{J_t} \frac{\partial F_a}{\partial \beta} & 0 & \frac{1}{J_t}
    \end{bmatrix} \boldsymbol{u}. \nonumber
\end{align}
Notably, the first column of the system matrix $\boldsymbol{A}$ contains only
zeros, so the matrix has an eigenvalue at $s=0$ and is not asymptotically stable
in $\theta$. However, with the definition of $\tilde{\theta}$ given above, all
perturbation modes of the linearized system are feedback-stabilizable. The last
two columns of the input matrix $\boldsymbol{B}$ are also of interest, showing
that the $\tilde{\tau}_g$ and $\tilde{\tau}_p$ inputs control the generator and
platform dynamics, respectively, in an uncoupled manner without directly
interfering with other DOFs. The blade pitch input has a direct impact on both
DOFs that cannot be decoupled.

The linear model suffers from the presence of NMPZs under conditions that are
common during operation. In~\cite{fischer2013}, the FOWT system is shown to be
non-minimum phase when
\begin{equation} \label{eq:nmpz}
    h_t^2 \left( \frac{\partial F_a}{\partial v} - \frac{\partial \tau_a}{\partial v}
        \frac{\partial F_a / \partial \beta}{\partial \tau_a / \partial \beta} \right)
    < -D_t.
\end{equation}
Within the wind turbine operating regions, the partial derivatives with respect
to $v$ are always positive, and those with respect to $\beta$ are always
negative. For USFLOWT, inequality~\eqref{eq:nmpz} is satisfied at operating
points just above rated wind speed and not satisfied for very high wind speeds,
though the precise operating point above which NMPZs are no longer present is
difficult to estimate due to the uncertain accuracy of the damping term in the
empirical platform pitch model.  When NMPZs are present, even relatively modest
gains can lead to system instability~\cite{larsen2007}. Detuning the controller
to avoid instability, described in Section~\ref{sec:detuning}, circumvents the
problem at the cost of generator speed tracking performance. Techniques for
addressing the NMPZ problem directly are described in Section~\ref{sec:parallel_comp}.

\section{CONTROL DESIGN} \label{sec:control_design}

\begin{figure}[t]
    \centering
    \includegraphics[clip, width=0.45\textwidth]{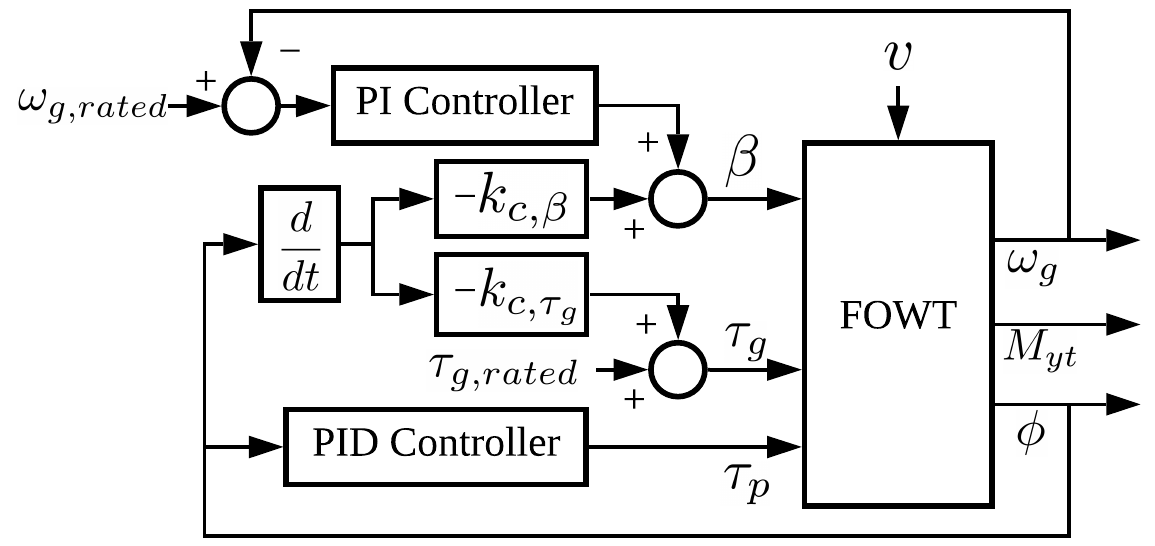}
    \caption{Block diagram of USFLOWT controller components.}
    \label{fig:controller_bd}
    \vspace*{-0.1cm}
\end{figure}

The USFLOWT controller is adapted from the reference open-source controller
(ROSCO) and is tuned using an adaptation of the procedure implemented in the
ROSCO toolbox~\cite{abbas2021}. The above-rated blade pitch controller uses
proportional-integral (PI) feedback to regulate generator speed and uses a
model-based automated tuning procedure to try to meet a desired closed-loop
transient response. Further, ROSCO has a built-in parallel compensation signal
that feeds back tower-top pitch rate to blade pitch through a proportional gain,
adding to the blade pitch control signal computed by the PI controller.
Additional FOWT-specific feedback loops considered in this study were
implemented on top of a baseline ROSCO controller using a Simulink~\cite{matlab}
interface to OpenFAST~\cite{openfast}. A simplified block diagram of the
complete controller is shown in Fig.~\ref{fig:controller_bd}.

The control gains are automatically tuned using the low-order linear model of
the FOWT system presented in Section~\ref{sec:wind_turbine}. Each control
component is tuned without considering the impact on other components or other
dynamics of the system except those captured by the tuning model. It is expected
that performance could be improved by using a coupled tuning process or
iterative closed-loop optimization of multiple parameters~\cite{zalkind2020},
though applying such a process to USFLOWT is left for a future study.

\subsection{Baseline Fixed-Bottom Wind Turbine Controller}

The baseline controller tested for USFLOWT was designed without accounting for
floating effects. Land-based wind turbines commonly use a PI controller to drive
pitch actuators on each blade to regulate the generator speed to its rated
value, and this is the function of the baseline controller implemented in ROSCO.
Utilizing the generator speed integral error in feedback raises the order of the
system in Eq.~\eqref{eq:rotor_lin}. The second-order transient response can be
tuned to satisfy a desired closed-loop natural frequency $\omega_{pi}$ and
damping $\zeta_{pi}$. At a given operating point, the second-order linear model
is used to calculate PI gains $k_p$ and $k_i$ that achieve the desired
closed-loop response:
\begin{equation} \label{eq:pi_gains}
    k_p = \frac{\frac{N_g}{J_r} \frac{\partial \tau_a}{\partial \omega_g} + 2 \zeta_{pi} \omega_{pi}}
    {\frac{N_g}{J_r} \frac{\partial \tau_a}{\partial \beta}},\ 
    k_i = \frac{\omega_{pi}^2}
    {\frac{N_g}{J_r} \frac{\partial \tau_a}{\partial \beta}}
\end{equation}
Because these gain values are in general unique to each linear model, they are
used to construct a piecewise gain schedule indexed for each operating point of
interest. The ROSCO implementation uses interpolated indexing of
$\overline{\beta}$ to determine the current operating point, which approximates
the current operating point without estimating the wind speed~\cite{abbas2021}.
Though this automated tuning process neglects many higher-order and nonlinear
effects, it is a suitable method for designing the bandwidth of a controller
without iterating on results from costly nonlinear simulations. 

For the baseline controller, the generator torque is fixed at its rated value in
above-rated winds.  When turbulence takes the wind turbine into below-rated wind
speeds, the blade pitch saturates and the generator torque is controlled using a
$k\omega^2$ law~\cite{abbas2021,pao2011}.

\subsection{Detuning} \label{sec:detuning}

A FOWT controller designed without accounting for floating effects is often
unstable in operating regions with NMPZs because of high feedback gains and a
closed-loop bandwidth that excites the platform pitch
resonance~\cite{jonkman2008}. In the worst case, this can cause the platform to
tip over, but often it will instead oscillate between nearby operating points:
unable to settle due to instability but kept from exponential growth by
nonlinear dynamics. While not catastrophic, the latter behavior is still
undesirable.

One method to avoid exciting the instability is to tune the closed-loop
bandwidth of the blade pitch controller to be considerably lower than the
platform natural frequency~\cite{larsen2007}.  Within the ROSCO automated tuning
procedure, the simplest approach to detuning is to heuristically reduce the
desired natural frequency $\omega_{pi}$. This is no more complex than tuning the
baseline controller and has some performance improvement where the instability
is worst, but the detuned controller expectedly has a degradation in tracking
performance at the operating points where detuning is used. A more sophisticated
approach uses the linear model to estimate a stability margin and tunes the
controller to the fastest response possible while maintaining the stability
margin above some threshold, as in~\cite{lemmer2020}. Accurately estimating the
stability margin to construct a tuning schedule requires a model more accurate
than the one used here, so the method of detuning in this work is based on a
simple linear automated tuning schedule that interpolates between a low
bandwidth at wind speeds just above rated and a higher bandwidth at higher wind
speeds where the system is more naturally stable. The upper and lower bounds, as
well as the shape of the schedule curve, are not the focus of this work; instead
they only serve as a starting point to take advantage of improved performance of
more advanced FOWT control methods. There is potential to improve the controller
performance by optimizing these and other design parameters in a robust
model-based detuning~\cite{lemmer2020} or an iterative co-design
process~\cite{zalkind2020}.

\subsection{Parallel Compensation} \label{sec:parallel_comp}

A common FOWT-specific control feature is an additional control loop that uses
feedback of rotor fore-aft velocity, measured at the nacelle.  This so-called
parallel compensation is a step towards MIMO control, designed using parallel
loop closure and tuned using the same low-order model as the baseline
controller. This method attempts to reduce the coupling between the competing
aerodynamics of rotor torque and thrust that come about while regulating
generator speed through blade pitch.  Parallel compensation has been shown to
be effective when the compensating actuator is the blade
pitch~\cite{abbas2021,fleming2016,vanderveen2012,lemmer2020} or the generator
torque~\cite{fischer2013,yu2018}. 

In this study, the fore-aft velocity signal used for parallel compensation is
the tower-top pitch rate, which is identical to platform pitch rate if the
tower is assumed to be rigid. However, when the control is implemented on a
flexible tower, the tower-top pitch motion directly impacts the response of the
rotor and is impacted by actuators. In the parallel compensation control
implementation, tower-top pitch rate is tapped directly from the computed
signal in OpenFAST~\cite{openfast}, with some filtering. The sensor model is
not considered in this work, but it is possible to reconstruct the tower-top
velocity from physical measurements.

Both blade pitch and generator torque are considered for parallel compensation
in this work, and a combination of both is shown to be a good compromise between
the benefits and drawbacks of each. The ROSCO controller implements parallel
compensation using blade pitch and offers an automatic tuning
method~\cite{abbas2021} which serves as a starting point for the tuning method
used in this work. The same linear model described in Section~\ref{sec:modeling}
is also used to automatically tune the parallel compensation gain. Considering
Eq.~\eqref{eq:ss_model}, the term $\boldsymbol{A}(2,4)$ captures the effect of
fore-aft motion on generator speed change. If this term were 0, the platform
pitching would have little direct effect on generator speed tracking. By this
tuning method, the parallel compensation feedback does not directly reduce the
platform motion, but instead compensates for the effect that the platform motion
has on generator speed regulation, and in turn increases the overall closed-loop
system stability. 

\subsubsection{Blade Pitch}

Parallel compensation using blade pitch feedback is built into ROSCO, but the
implementation was modified slightly for this study. While ROSCO uses a single
gain for the fore-aft velocity feedback term, we use a gain schedule that keeps
the parallel compensation gain consistent with the PI controller gain for each
operating point.  The blade pitch compensation, $\beta_c$, uses proportional
feedback of the platform pitch rate, which can also be thought of as derivative
feedback of the platform pitch angle:
\begin{equation} \label{eq:beta_c}
    \beta_c = -k_{c,\beta} \dot{\phi}
\end{equation}
This term is added to the blade pitch command generated by the PI controller
before the actuator saturation limits are applied.  In the state-space model of
Eq.~\eqref{eq:ss_model}, this feedbacks subtracts the term $k_{c,\beta}
\frac{N_g}{J_r} \frac{\partial \tau_a}{\partial \beta}$ from
$\boldsymbol{A}(2,4)$. Choosing a gain that sets $\boldsymbol{A}(2,4)=0$ fully
compensates the effect of platform pitch on generator speed, however, due to
the coupling of blade pitch to both aerodynamic torque and thrust, such a gain
also considerably reduces the effective system fore-aft damping as a side
effect. It is therefore desirable to choose a smaller gain to partially
compensate the fore-aft motion.  The full-compensation gain for blade pitch is
\begin{equation} 
    \gamma_{c,\beta} = -h_t \frac{\partial \tau_a}{\partial v}
    \left( \frac{\partial \tau_a}{\partial \beta} \right)^{-1},
\end{equation}
and setting $k_{c,\beta} = m_\beta \gamma_{c,\beta}$, $m_\beta \in [0, 1]$
allows the control system designer to select the degree of partial compensation
from the blade pitch actuator. The fully-compensated tuning is performed by the
ROSCO toolbox~\cite{abbas2021} with $m_\beta = 1$, but ROSCO uses additional
filtering to change the dynamics of the feedback loop even further thereby
effectively reducing $m_\beta$ at certain frequencies. Note that
$\gamma_{c,\beta} > 0$, so the $\beta_c$ contribution is positive for a forward
swinging rotor ($\dot{\phi} < 0$).

Per the analysis in~\cite{fischer2013}, this feedback redistributes the impact
of NMPZs, but it does not remove them. Algebraically, the same term is added to
both sides of inequality~\eqref{eq:nmpz}, so the condition is still satisfied
regardless of the selected gain. As demonstrated in~\cite{yu2018}, allowing
$m_\beta~<~0$ is an alternative tuning method that attempts to increase
fore-aft damping at the expense of generator speed tracking.

\subsubsection{Generator Torque}

Generator torque feedback is in theory more suitable for compensating the effect
of platform motion on generator speed. Unlike the blade pitch, generator torque
feedback is capable of modifying the generator dynamics without impacting the
fore-aft damping as a side effect and can therefore mitigate or completely
remove the NMPZs~\cite{fischer2013,yu2018}. Following the same tuning approach:
\begin{align} 
    \tau_{g,c} &= -k_{c,\tau_g} \dot{\phi} \nonumber \\
    \gamma_{c,\tau_g} &= \frac{h_t}{N_g} \frac{\partial \tau_a}{\partial v} \\
    k_{c,\tau_g} &= m_{\tau_g} \gamma_{c,\tau_g},\ m_{\tau_g} \in [0,1] \nonumber 
\end{align}
One limitation to using generator torque for parallel compensation is a strict
limit on the maximum resistance torque that can be supplied by the actuator.  In
this study, the generator torque maximum $\tau_{g,max}$ is set to 120\% of
$\tau_{g,rated}$. Using the full-compensation gain makes the system minimum
phase for all operating points, however, in practice, saturation of $\tau_g$
prohibits actuator signals large enough to achieve full compensation. As before,
it is beneficial to reduce the gain with $m_{\tau_g} < 1$ to avoid introducing
unexpected behavior at the saturation limit while compensating typical fore-aft
velocities experienced during operation. This parallel compensation using
$\tau_g$ feedback changes inequality~\eqref{eq:nmpz} to
\begin{equation} \label{eq:nmpz_comp}
    h_t^2 \left( \frac{\partial F_a}{\partial v} - (1 - m_{\tau_g}) \frac{\partial \tau_a}{\partial v}
        \frac{\partial F_a / \partial \beta}{\partial \tau_a / \partial \beta} \right)
    < -D_t.
\end{equation}
With full compensation ($m_{\tau_g} = 1$), inequality~\eqref{eq:nmpz_comp} is no
longer satisfied because the term on the left is positive, demonstrating that
the NMPZs are not present. With low enough $m_{\tau_g}$, the NMPZs are partially
compensated and may still be present for certain operating points if the natural
damping $D_t$ is small.  If the maximum expected platform pitch rate at an
operating point is $\dot{\phi}_{max}$, then the degree of partial compensation
can be set to
\begin{equation} \label{eq:m_tau}
    m_{\tau_g} = \frac{\tau_{g,max} - \tau_{g,rated}}{-\gamma_{c,\tau_g} \dot{\phi}_{max}}
\end{equation}
so that the computed $\tau_g$ compensating for a pitch rate $\dot{\phi}_{max}$
is at most $\tau_{g,max}$. An additional benefit to limiting the gain is the
reduction of drivetrain loads, as pointed out in~\cite{fischer2013} (where the
authors use a constant $m_{\tau_g}=0.5$). Drivetrain loads are not analyzed in
the present study.

\subsubsection{Dual Compensation}

Given the particular tradeoffs between blade pitch and generator torque as used
for parallel compensation, it is sensible to attempt to combine them in a dual
parallel compensation control scheme. Using partial compensation of both blade
pitch and generator torque, the original objective of setting
$\boldsymbol{A}(2,4)=0$ in~\eqref{eq:ss_model} will be satisfied if $m_\beta +
m_{\tau_g} = 1$, so the share of effort managed by each actuator can be tuned.
In this work, the dual compensation is tuned at each operating point using the
generator torque saturation as a constraint (using Eq.~\eqref{eq:m_tau}), and
setting $m_\beta~=~1 - m_{\tau_g} < 1$ achieves full compensation.

\subsection{Platform Control} \label{sec:ptfm_control}

Two platform actuation options exist for the SpiderFLOAT: actuators that operate
at a low bandwidth and a high bandwidth. Both options are being investigated for
their benefits to power generation and structural loads mitigation, and the
realization of either depends on existing technology for reasonable cost.

\begin{figure*}[t]
    \centering
    \includegraphics[clip, trim=0cm 0cm 0cm 0cm, width=1.00\textwidth]{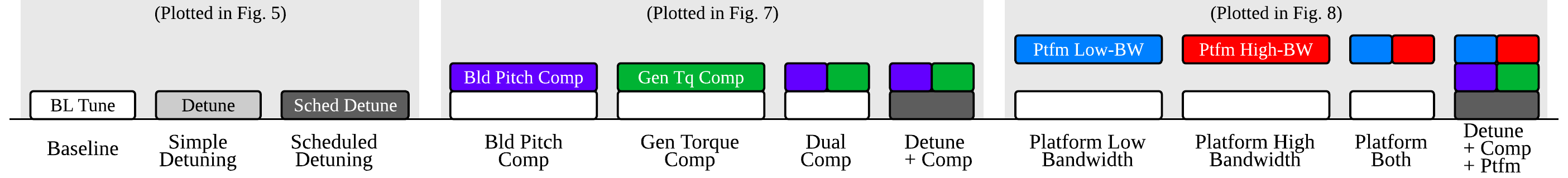}
    \vspace*{-.75cm}
    \caption{Components of each controller, where each column-stack is a single
    controller evaluated in this paper. Each row represents one control feature:
    the bottom row is basic PI tuning method, the middle row is parallel
    compensation, and the top row is platform control. Each group of columns
    outlined by a light gray rectangle shares a performance plot in the sections
    below.  Component color is consistent for each added component and is only
    labeled once. For example, the ``Detune + Comp + Ptfm" controller uses
    scheduled detuning, both parallel compensation actuators, and both platform
    actuators. Due to the large number of controllers in this paper, the colors
    in this figure are not always consistent with the colors used in performance
    plots.}
    \label{fig:controller_table}
    \vspace*{-0.2cm}
\end{figure*}

\subsubsection{Low-Bandwidth}

Using variable-ballast buoyancy cans, the SpiderFLOAT legs can be selectively
weighted to influence the quasi-static settling heel angle.  Changing the
ballast in the cans can be achieved using air compressors and/or water pumps to
fill or deplete the cans of seawater. These actuators would operate at a low
bandwidth, with settling time on the order of several minutes, similar to a
conventional wind turbine yaw servo~\cite{pao2011}. Our design for controlling
this actuator is a slow integral controller compensating a slowly moving average
heel angle.  Additionally, it may be possible to utilize marine forecast data in
a low-bandwidth feedforward controller to anticipate trends in mean wind speed
and direction corresponding to platform heel angle.

\subsubsection{High-Bandwidth}

The high-bandwidth platform actuation option would dynamically reel in or out
stay cables connecting the SpiderFLOAT legs to the central stem
(Fig.~\ref{fig:sf}). These cables are under enormous tension even in a static
condition, so changing their length by an appreciable amount would require
considerable force, akin to a crane lifting building materials weighing
kilotons. The maximum possible bandwidth of a physically realizable actuator is
still being investigated, but for the purposes of this work the actuator is
assumed to operate faster than the platform pitch mode, with an actuation time
faster than 15 seconds. This actuator model is presented as the potential
best-case platform control capability to contrast with the lower-risk can
ballast actuator.

A high-bandwidth platform controller is implemented using a
proportional-integral-derivative (PID) controller designed with a closed-loop
bandwidth above the platform pitch natural frequency in order to actively
dampen platform motion. Utilizing the second-order platform pitch model in
Eq.~\eqref{eq:ptfm}, the PID controller was tuned for a desired closed-loop
bandwidth using classical linear control design methods, rather than a fully
automated procedure used in the earlier tunings discussed.  An analytical
tuning method accounting for the SpiderFLOAT dynamics is pending development of
a higher-order control-oriented linear platform model.

\section{SIMULATION RESULTS} \label{sec:simulation_results}

\begin{figure}[t]
    \centering
    \includegraphics[width=0.48\textwidth]{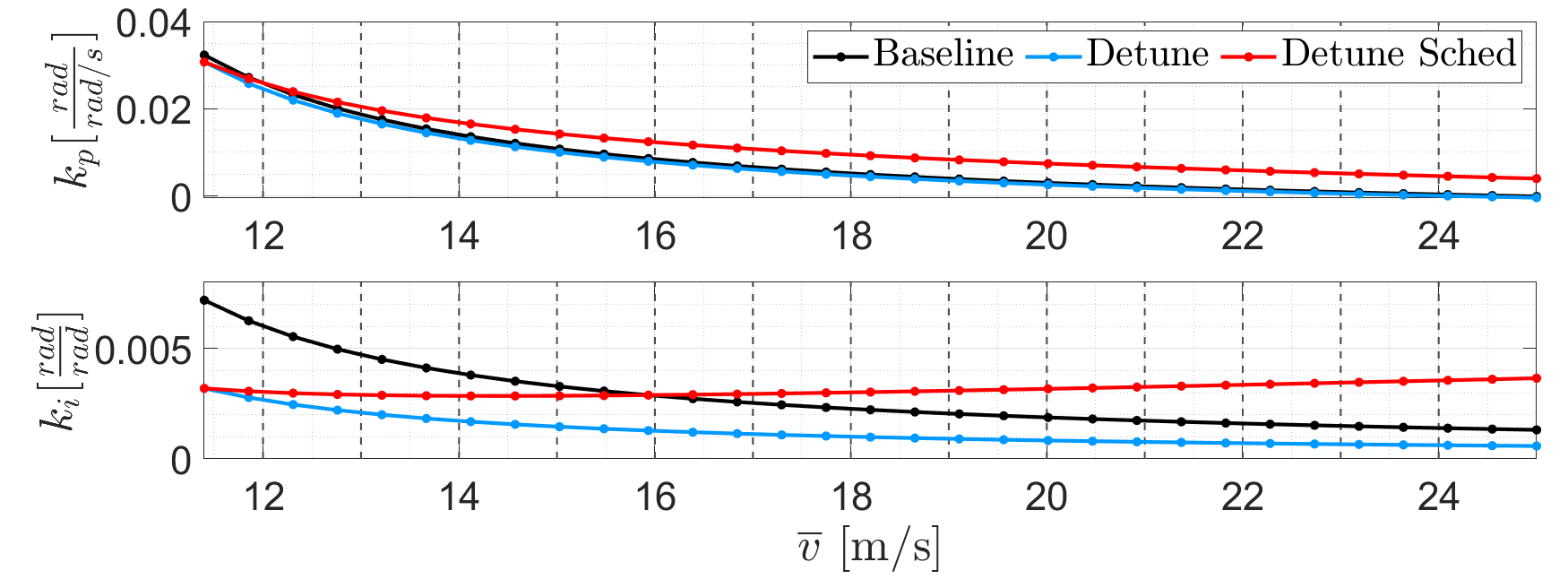}
    \vspace*{-.75cm}
    \caption{PI gains for baseline and detuned controllers. Wind speed operating
    points used for tuning (dots) have finer granularity than
    wind speed references used as simulation data points (vertical dashed
    lines).}
    \label{fig:detuning}
    \vspace*{-0.1cm}
\end{figure}

Each controller presented has been simulated for the USFLOWT system using the
open-source aero-hydro-servo-elastic wind turbine simulation tool
OpenFAST~\cite{openfast}. The system was simulated with 6 random turbulence
seeds at each whole-numbered wind speed between 12 and 24 m/s (for a total of 78
simulation cases for each controller) using the normal turbulence model (NTM) as
outlined in design load case 1.1 of the International Electrical Commission
(IEC) standard 61400-1~\cite{iec}. At each reference wind speed, corresponding
wave conditions are informed by metocean data for a site near Monhegan Island,
Maine.  While a realistic wave disturbance was simulated in every case, it has
not been found to have a significant impact on USFLOWT performance, so wave
effects have not been factored into the control design presented here.  All
simulations used a duration of 800 seconds, with the first 200 seconds of
transient settling discarded. A summary of the components of each controller
simulated can be found in Fig.~\ref{fig:controller_table}.

The system outputs used as performance metrics to compare controllers are
generator speed and tower base fore-aft bending moment, with performance
measured by the standard deviation and absolute maximum quantities of each.
Large generator overspeed spikes, exceeding rated speed by approximately 20\%,
can cause a safety shutdown of the wind turbine that would lead to a loss in
power production, while the standard deviation (STD) measures tracking
performance in the presence of random variation, i.e. turbulent simulation.
Standard deviation of the tower bending moment approximately correlates with
fatigue load, while the maximum correlates with extreme load.  For both metrics,
lower values signify better performance. In all results shown, the generator
speed deviation is normalized to its rated value, and the tower base moment is
normalized to the mean moment experienced by the baseline controller at 12 m/s.

\subsection{Detuning}

The baseline controller was tuned for a closed-loop bandwidth and damping of
$\omega_{pi} = 0.3$ rad/s and $\zeta_{pi} = 0.7$.  The simple detuned controller
was designed for a slower response, with a bandwidth and damping of $\omega_{pi}
= 0.2$ rad/s and $\zeta_{pi} = 1.0$, similar to the values used
in~\cite{lemmer2020}.  When this detuning was shown to improve generator speed
tracking performance at low wind speeds but reduce performance at high wind
speeds, a scheduled detuning approach was employed that tuned the bandwidth and
damping to slower values at low winds and faster values at high winds,
interpolated linearly.  The low-speed transient response was set to the simple
detuned values, and the high-speed controller was tuned with $\omega_{pi} = 0.5$
rad/s and $\zeta_{pi} = 0.7$. 

\begin{figure}[t]
    \centering
    \includegraphics[width=0.48\textwidth]{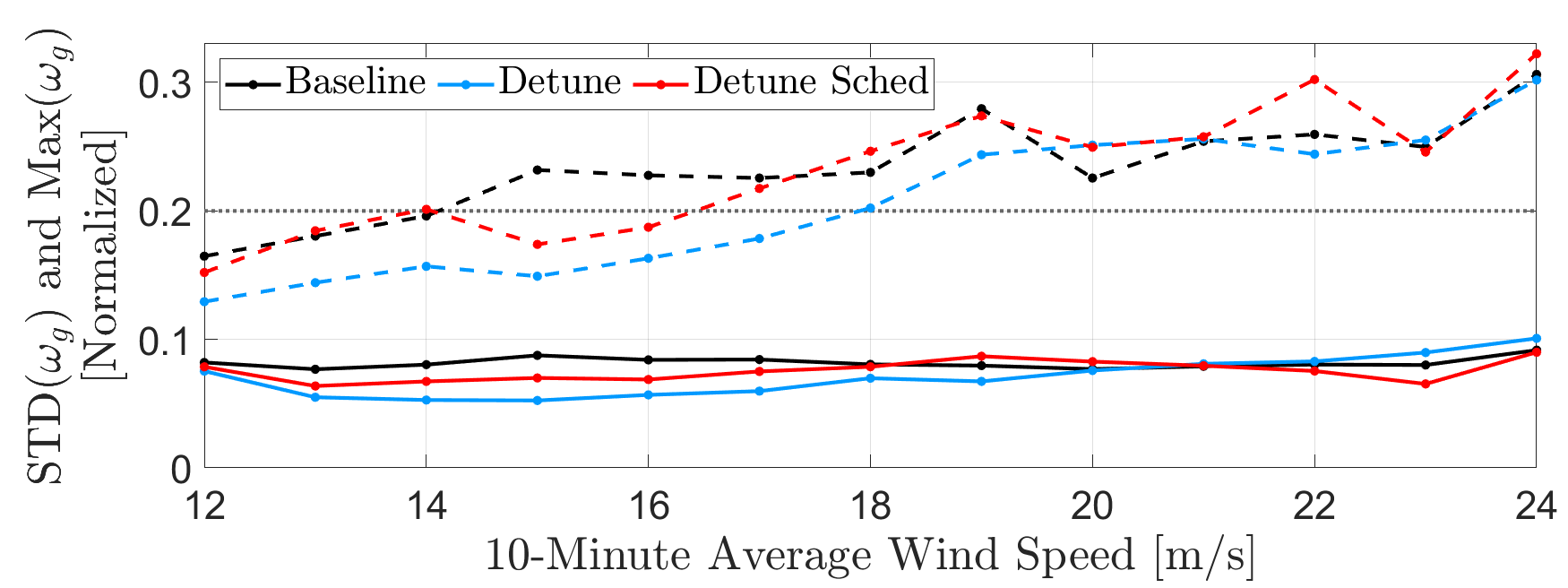}
    \vspace*{-.75cm}
    \caption{Simulated generator speed standard deviation (solid) and maximum
    (dashed) for detuned controllers. Reference wind speed data points are shown
    by dots. The gray dotted line is the maximum safety threshold.}
    \label{fig:detuning_perf}
    \vspace*{-0.1cm}
\end{figure}

The PI gains of the three controllers are plotted in Fig.~\ref{fig:detuning}.
The simulated performance of the three controllers is compared in
Fig.~\ref{fig:detuning_perf}. The generator speed tracking performance at lower
above-rated wind speeds is improved slightly by both detuned controllers, and
though not shown, the mean generator speed is also higher, which translates to
larger mean power during operation. At higher wind speeds, all controllers fail
to stop transient spikes that exceed rated generator speed up to 30\%, even
though the mean wind speed is outside of the domain where the system has NMPZs.
If deployed in the field, a generator overspeed of this magnitude would likely
cause the turbine to shut down. The scheduled detuned controller does not
noticeably increase performance beyond either the baseline or fully-detuned
controller without advanced FOWT control features, but its faster bandwidth at
higher windspeeds contributes to better tracking when parallel compensation is
also employed in the next section.

\subsection{Parallel Compensation}

For the standalone blade pitch parallel compensation case, it was found that
using full compensation feedback with $m_\beta = 1$ decreased fore-aft damping
enough to be detrimental to system performance. Thus, the gain was set so that
$m_\beta~=~0.5$ as a tradeoff between generator speed tracking and fore-aft
damping.  While this approach is simple and shows some improvement over the
baseline, it could be improved further by a more focused optimization.

\begin{figure}[t]
    \centering
    \includegraphics[width=0.48\textwidth]{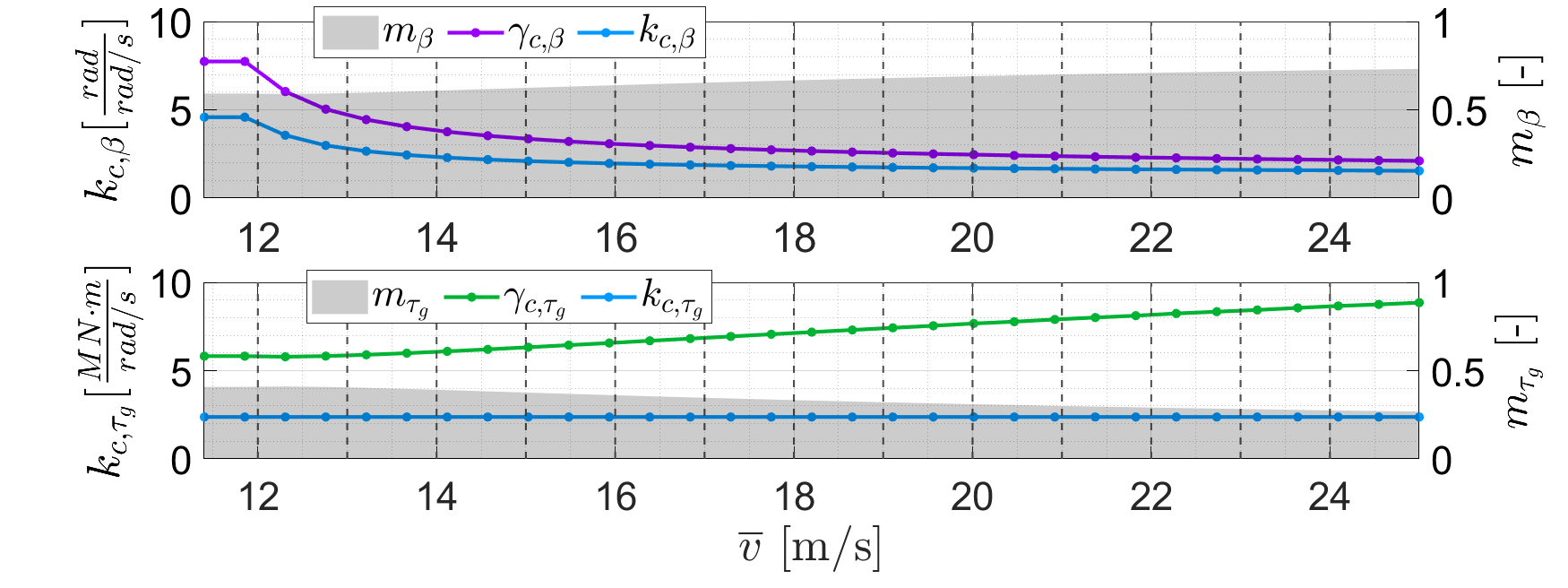}
    \vspace*{-.75cm}
    \caption{Parallel compensation gains for the dual compensation controller.
    The shaded region $m_x$ is the fraction of full compensation effort for each
    actuator used by the dual compensation tuning. Note that the $m_\beta~+~m_{\tau_g}~=~1$ and $k_{c,x} = m_x \gamma_{c,x}$, where $x$ is $\beta$ or
    $\tau_g$.}
    \label{fig:pc_gains}
    \vspace*{-0.1cm}
\end{figure}

The generator torque parallel compensation gain was tuned to avoid saturation as
described in Section~\ref{sec:parallel_comp}. During simulations using a
controller without parallel compensation, the maximum platform pitch rate was
observed to be typically near $\dot{\phi} = 1^\circ/s = 0.0175$ rad/s.  From
Eq.~\eqref{eq:m_tau}, this limit implies a gain $k_{c,\tau_g}$ that is held
constant across all operating points.  Because $\gamma_{c,\tau_g}$ is larger for
higher winds, maintaining a constant gain requires decreasing $m_{\tau_g}$ for
operating points at high wind speeds, and therefore the generator torque
feedback compensates less of the total fore-aft motion. When the actuators are
combined for dual parallel compensation, the generator torque tuning remains the
same, while the blade pitch controller is tuned to satisfy the remainder of the
full compensation effort.  At high wind speeds, the decrease in $m_{\tau_g}$
requires increasing $m_\beta$, leading to a more aggressive blade pitch
controller where the side effect from blade pitch feedback is less severe. The
parallel compensation gain schedules for $\beta$ and $\tau_g$ are illustrated in
Fig.~\ref{fig:pc_gains}. 

\begin{figure}[t]
    \centering
    \includegraphics[width=0.48\textwidth]{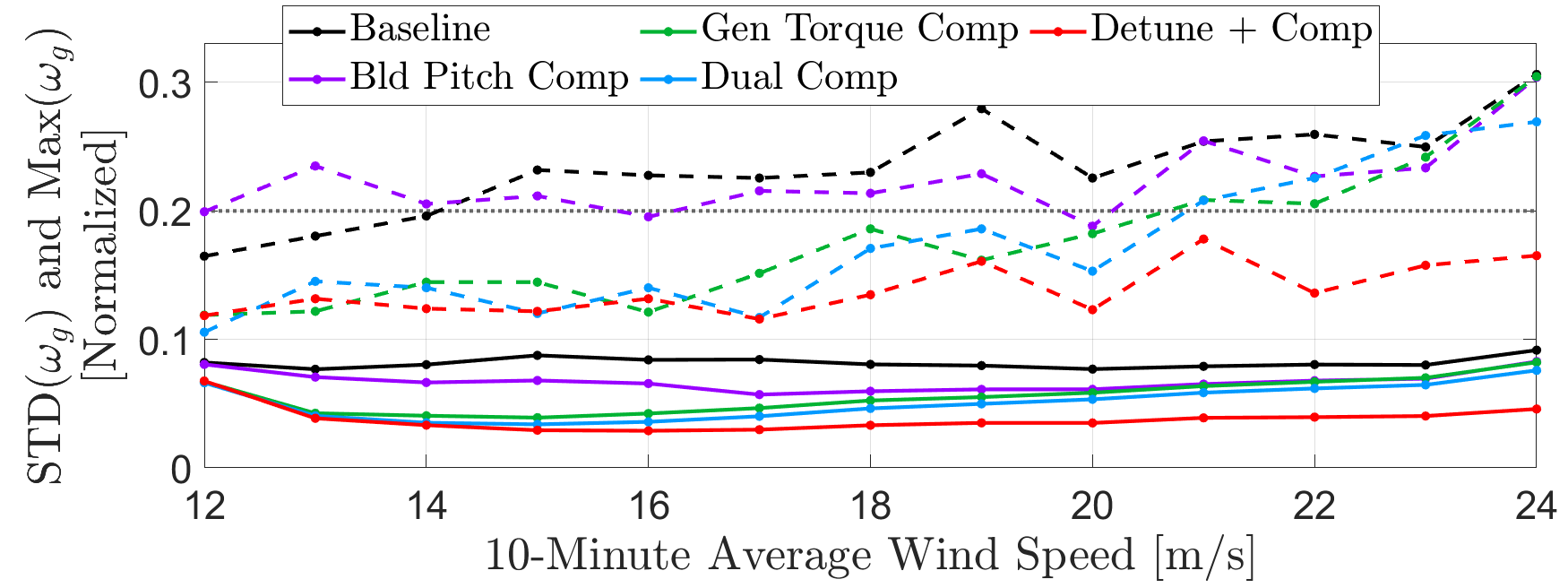}
    \vspace*{-.75cm}
    \caption{Simulated generator speed standard deviation (solid) and maximum
    (dashed) for parallel compensation controllers. The ``Detune + Comp"
    controller uses scheduled detuning and dual parallel compensation.}
    \label{fig:pc_perf}
    \vspace*{-0.1cm}
\end{figure}

The generator speed tracking performance for parallel compensation techniques is
compared against the baseline in Fig.~\ref{fig:pc_perf}. Blade pitch
compensation shows modest improvement of STD over the baseline, while sometimes
exacerbating maximum overspeed spikes, especially just above rated wind, where
the thrust side effect is highest. In contrast, the generator torque
compensation significantly improves tracking and reduces maximum spikes just
above rated, but the improvement diminishes to match that of the blade pitch at
higher winds. The dual compensation approach matches the generator torque-only
compensator just above rated, but improves modestly over either individual
compensator at higher winds. Only after combining the detuning schedule from
Section~\ref{sec:detuning} does the addition of parallel compensation show
significant improvement over the baseline at higher winds. In fact, the
controller with both scheduled detuning and dual compensation does not exceed a
20\% safety threshold above rated generator speed for any test case.

The complementary performance regions of both parallel compensation schemes
suggests a more ideal dual compensation tradeoff that could improve performance
beyond that shown here. Tuning the blade pitch compensation gain $k_{c,\beta}$
with the opposite sign (as in~\cite{yu2018}) at lower wind speeds would decrease
the tracking performance but increase nacelle fore-aft damping. Since the risk
of generator overspeed events is not as severe at lower wind speeds, the damping
benefit of such a tuning approach would be advantageous paired with the partial
compensation from generator torque.  Scheduling the compensation tradeoff to
increase fore-aft damping just above rated winds while keeping the good
compensation performance at higher winds is the next step for the control design
on USFLOWT.

\subsection{Platform Control}

The OpenFAST simulation tool currently does not allow controlling the buoyancy
of floating elements in closed-loop, so the low-bandwidth actuator was simulated
by changing fixed model parameters a-priori for each simulation. This is an
effective time-representation of how the actuator would behave on a physical
system due to the slow actuation speed, on the order of the duration of an
entire simulation.

The high-bandwidth actuator was implemented in closed-loop using SubDyn cable
control in OpenFAST through a Simulink interface. The PID cable controller was
tuned for a closed-loop bandwidth of 1 rad/s = 0.16 Hz (between the platform and
tower natural frequencies) using Matlab's Control Systems Designer
toolbox~\cite{matlab}.

\addtolength{\textheight}{-2.2cm}   

\begin{figure}[t]
    \centering
    \includegraphics[width=0.48\textwidth]{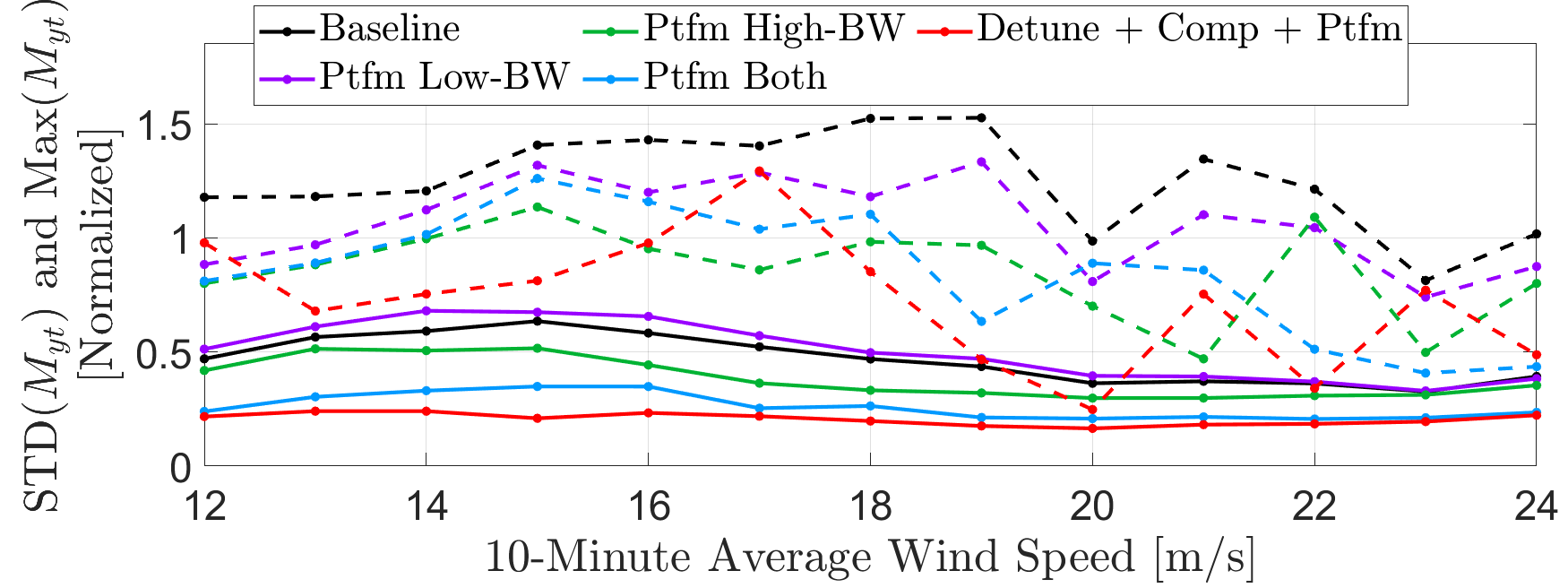}
    \vspace*{-.75cm}
    \caption{Simulated tower base fore-aft bending moment standard deviation
    (solid) and maximum (dashed) for platform controllers. The ``Detune + Comp +
    Ptfm" controller uses scheduled detuning, dual parallel compensation, and
    both platform actuators.}
    \label{fig:ptfm_load_perf}
    \vspace*{-0.1cm}
\end{figure}

The benefits of this application of platform control are not primarily to
generator speed but to tower loading, captured by the tower base fore-aft
bending moment, $M_{yt}$. The effects on tower loading of the platform
controllers are compared in Fig.~\ref{fig:ptfm_load_perf}.  When combined with
the best controller from previous cases that includes scheduled detuning and
dual parallel compensation, the performance improves further.

The tower base fore-aft bending moment is highly correlated to platform pitch
motion. Significant platform pitch deviation leads to large tower fatigue loads,
and extreme-pitch events (large deviations from the mean) also translate to
spikes in the extreme tower load. Therefore, targeting a reduction of platform
pitch deviation and mean also reduces tower fatigue and extreme loading,
respectively. A controller that improves structural loading allows for reduction
of LCOE primarily through the redesign of structural components with smaller
load margins, thereby reducing capital expenditures. This multi-disciplinary
co-design process used during USFLOWT development enables LCOE reduction through
iterative structural redesign in tandem with control improvements, leading to a
final design that is more optimal than designing each aspect of the system
individually~\cite{garcia-sanz2019}.

\begin{figure}[t]
    \centering
    \includegraphics[width=0.48\textwidth]{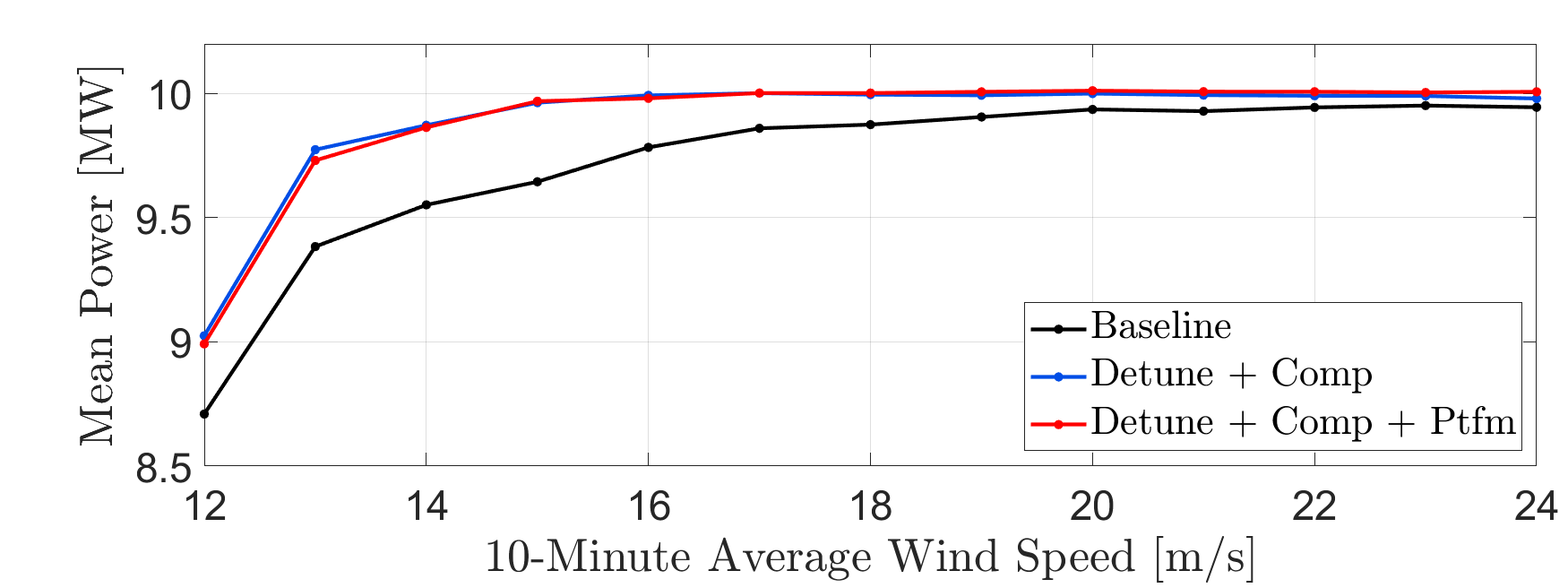}
    \caption{Simulated above-rated mean power for baseline and combined
    controllers. See Fig.~\ref{fig:controller_table} for a description of
    the combined controllers.}
    \label{fig:power_perf}
    \vspace*{-0.1cm}
\end{figure}

A more direct means for a controller to reduce LCOE is through an increase in
mean power production. The power output is a product of the generator speed and
generator torque, so it is not captured well by the linearized model and is
therefore not used directly as an objective for the control design. It is still
valuable to examine differences in mean power across controllers since mean
power can suffer from the FOWT instabilities previously discussed. The mean
power for the combined controllers of Sections~\ref{sec:parallel_comp}
and~\ref{sec:ptfm_control} is shown in Fig.~\ref{fig:power_perf}, where there is
a difference of several percent between the baseline and improved FOWT
controllers. Because the costs of a wind turbine are typically fixed after
deployment, an increase in power output without increased structural loads
translates directly to profit, so power increases of even a few percent are
beneficial to the industry and overall adoption of the technology. 

\section{CONCLUSIONS} \label{sec:conclusions}

Several controller techniques for improving generator speed tracking and
reducing tower loads have been presented with their results in application to
the USFLOWT system. FOWT control features supported by the literature have been
shown to improve generator speed tracking and power generation when applied to
USFLOWT. Additionally, novel platform actuators allow the USFLOWT controller to
apply lesser studied control techniques that show promise for reducing
structural loads. The design and control of these platform actuators will be
considered as parameters in future USFLOWT multi-disciplinary design
optimization, adding more dimensions to the design space which may allow further
reductions in LCOE, as long as actuator manufacturing and operational costs are
not too high. 

Co-design of the controller and platform structural parameters offers promise
for yielding a more optimal solution than the traditional sequential design
process, whereby the wind turbine is designed first followed by the control
design.  Indeed, the controller application shown here is only a starting point
for a more sophisticated co-optimization that takes into consideration
higher-order effects and will offer solutions connecting varied design aspects.
USFLOWT aims to explore optimal regions of the design space while furthering the
state of the art in floating wind energy.

\end{document}